\documentclass[english,12pt]{article}
\usepackage{array}
 \usepackage[normalem]{ulem}
\usepackage{graphicx}
\usepackage{amssymb}
\usepackage[english]{babel}
\usepackage{amsmath}
\usepackage{braket}
\usepackage{multirow}
\usepackage{prettyref}
\usepackage{babel}
\usepackage{units}
\usepackage[latin1]{inputenc}
\usepackage{amsfonts}
\usepackage{amssymb}
\usepackage{slashed}
\usepackage{babel}
\usepackage{color}
\usepackage[dvipsnames]{xcolor} 
\usepackage{cite}
\def\@fmsl@sh#1#2#3{\m@th\ooalign{$\hfil#1\mkern#2/\hfil$\crcr$#1#3$}}
 \def\eq#1\en{\begin{equation}#1\end{equation}}
\def\s[#1,#2]{[#1\stackrel{\star}{,}#2]}
\def\sx[#1,#2]{[#1\stackrel{\star_{x}}{,}#2]}

\usepackage[english]{babel}

\usepackage{setspace}

\usepackage{graphicx} 
\usepackage{graphics} 
\usepackage{float} 
\usepackage{subcaption} 

\usepackage{amsmath}
\usepackage{amssymb}
\usepackage{mathtools}
\usepackage{enumerate}
\usepackage{enumitem}
\usepackage[colorlinks=true, allcolors=black]{hyperref}

\usepackage{tensor}

\def\gsim{\mathrel{\rlap{\lower4pt\hbox{\hskip1pt$\sim$}}
		\raise1pt\hbox{$>$}}}       

\newcommand{\nc}{\newcommand}
\nc{\beq}{\begin{equation}}
\nc{\eeq}{\end{equation}}
\nc{\beqa}{\begin{eqnarray}}
\nc{\eeqa}{\end{eqnarray}}

\def\bc{\begin{center}}
\def\ec{\end{center}}

\def\gsim{\mathrel{\mathpalette\atversim>}}

\def\bc{\begin{center}}
\def\ec{\end{center}}

\def\gsim{\mathrel{\rlap{\lower4pt\hbox{\hskip1pt$\sim$}}

    \raise1pt\hbox{$>$}}}       

\def\gsim{\mathrel{\rlap{\lower4pt\hbox{\hskip1pt$\sim$}}
    \raise1pt\hbox{$>$}}}       

\begin{document}
\makeatletter
\def\fmslash{\@ifnextchar[{\fmsl@sh}{\fmsl@sh[0mu]}}
\def\fmsl@sh[#1]#2{%
  \mathchoice
    {\@fmsl@sh\displaystyle{#1}{#2}}%
    {\@fmsl@sh\textstyle{#1}{#2}}%
    {\@fmsl@sh\scriptstyle{#1}{#2}}%
    {\@fmsl@sh\scriptscriptstyle{#1}{#2}}}
\def\@fmsl@sh#1#2#3{\m@th\ooalign{$\hfil#1\mkern#2/\hfil$\crcr$#1#3$}}

\makeatother

\thispagestyle{empty}
\begin{titlepage}
\boldmath
\begin{center}
  \Large {\bf {Leading order effective operators in quantum gravity}}
    \end{center}
\unboldmath
\vspace{0.2cm}
\begin{center}
{\large Tommaso Antonelli}\footnote{E-mail: t.antonelli@sussex.ac.uk}$^a$,
{\large Xavier~Calmet}\footnote{E-mail: x.calmet@sussex.ac.uk}$^a$
{\large and}  {\large  Stephen D. H. Hsu}\footnote{E-mail: hsusteve@gmail.com}$^b$
 \end{center}
\begin{center}
$^a${\sl Department of Physics and Astronomy, \\
University of Sussex, Brighton, BN1 9QH, United Kingdom
}\\
$^b${\sl Department of Physics and Astronomy\\ Michigan State University, East Lansing, Michigan 48823, USA}\\
\end{center}
\vspace{5cm}
\begin{abstract}
\noindent
We explore the nature of higher dimensional operators generated by quantum gravity. Calculating the tree-level and one-loop effective operators generated by graviton exchange between fields of the standard model and those of a hidden sector, we show that the leading order operators generated by quantum gravity are non-local dimension 6 operators. Dimension 5 operators are not generated by perturbative (weak field) effects, although they might be generated by strong field effects such as Planck-scale fluctuations in spacetime. We investigate the consequences of our findings for models of ultralight scalar dark matter. 
\end{abstract}  
\end{titlepage}



\newpage

\section{Introduction}
In this paper we revisit an old question, namely that of higher-dimensional operators generated by quantum gravity. A folk theorem of modern theoretical physics is that the effective Lagrangian is a sum of all local operators allowed by the symmetries, with coefficients that depend on the underlying high-energy theory \cite{Weinberg:1996kr}. In other words, it is expected that all operators compatible with the symmetries of the model will be generated at some level by quantum effects. The exact source of this folk theorem is difficult to pinpoint but it is often attributed to Gell-Mann who emphasized that symmetries dictate interactions in the effective field theory approach. In his lectures (e.g., Caltech lectures from the 1970s, partially compiled in The Quark and the Jaguar \cite{Gell-Mann:1994aya}), he discussed how symmetries constrain the form of interactions, implying that all symmetry-allowed operators appear in the effective theory.

This folk theorem has been widely employed to discuss physics beyond the standard model and to estimate the magnitude of effects of new physics on observables. A textbook example is, e.g., the contribution of dimension five $ (1/\Lambda) \bar \psi \sigma_{\mu\nu} \psi F^{\mu\nu}$ or six $(m_\psi/\Lambda^2) \bar \psi \sigma_{\mu\nu} \psi F^{\mu\nu}$ operators to the anomalous magnetic moment of the muon see e.g. \cite{Buchmuller:1985jz}. 

When applied to quantum gravity, the very same folk theorem leads to the conclusion that quantum gravitational effects must connect the fields of the standard model to any fields beyond the standard model, generating operators of dimension five and higher depending on the spin of the fields in the new physics sector and the gauge symmetries in that new sector. For example, if there are singlet scalar fields $\phi$, operators of the type $(1/M_P) \phi F_{\mu\nu} F^{\mu\nu}$ should be generated, where $M_P$ is the reduced Planck mass and $F_{\mu\nu}$ the electromagnetic field tensor.

The aim of this work is to revisit this conclusion by deriving the leading-order operators generated by the tree-level exchange of a graviton as well as those generated at one-loop level. We consider the field content of the standard model and investigate how gravity couples these fields to those of a hidden sector. Our conclusions are that at tree level, the lowest-order operators generated are non-local dimension 6 ones while at the loop level dimension 8 operators are generated.  We consider implications of our findings for models with ultralight dark matter. Finally, we argue that perturbative quantum gravity does not generate higher-dimensional operators that break symmetries of the tree-level action.

\section{Higher dimensional operators generated by graviton exchange}
\label{sec:graviton_exchange}

The aim of this section is to derive the leading-order operators generated by the exchange of gravitons at tree-level. We consider a model constituting of scalar fields, fermions, and vector bosons which interact gravitationally and via a potential for the scalar field. We allow for a non-minimal gravitational coupling of the scalar field to the Ricci scalar. The action is given by 
\begin{equation}
  \label{eq:action_decomposition}
  S=S_{\rm EH}+ S_{\rm matter},
\end{equation}
where $S_{\rm EH}$ is the Einstein-Hilbert action
\begin{equation}
  S_{\rm EH}=-\frac{M_P^2}{2}\int d^4x\sqrt{-g}\ \mathcal{R}, \quad M_P=\frac{1}{\sqrt{8\pi G}},
\end{equation}
and where the matter part is given by
\begin{equation}
  \label{eq:matter_action}
  S_{\rm matter}=S_\phi+S_\psi+S_A,
\end{equation}
where $S_\phi$, $S_\psi$, $S_A$ are, respectively, the scalar field, fermion, and vector parts of the action.
The scalar field sector contains a minimal as well as a non-minimal coupling to the curvature:
\begin{equation}
  S_\phi=\int d^4x\sqrt{-g}\left(\frac{1}{2}g^{\mu\nu}\,\partial_\mu\phi\,\partial_\nu\phi-V(\phi)+\frac{1}{2} F(\phi)\,\mathcal{R}\right).
\end{equation}
We consider the standard Dirac action for the fermion field
\begin{equation}
  S_\psi=\int d^4x\sqrt{-g}\ \bar{\psi}\left(\frac{i}{2} \overset{\leftrightarrow}{\slashed{D}}-m_\psi\right)\psi,
\end{equation}
where the double arrow on top of the derivative operator denotes the following:
\begin{equation}
  f\,\overset{\leftrightarrow}{\partial}_{\mu}\,g= f\overset{\rightarrow}{\partial}_\mu g- f\overset{\leftarrow}{\partial}_\mu g.
\end{equation}
Here, $\slashed{D}$ denotes the covariant derivative with respect to the spin connection:
\begin{equation}
  \overset{\rightarrow}{\slashed{D}}\psi=\gamma^a\,e^\mu_a\left(\overset{\rightarrow}{\partial}_\mu+\frac{1}{2}\sigma^{bc}e^\nu_b\partial_\mu e_{c\nu}\right)\psi, \quad \bar {\psi}\overset{\leftarrow}{\slashed{D}}=\bar{\psi}\left(\overset{\leftarrow}{\partial}_\mu-\frac{1}{2}\sigma^{bc}e^\nu_b\partial_\mu e_{c\nu}\right)\gamma^a\,e^\mu_a,
\end{equation}
where $\sigma^{ab}=[\gamma^a,\gamma^b]/4$ and $e^{\mu}_a$ is the vielbein of the metric.\\
The electromagnetic vector field is described by the action
\begin{equation}
  S_A=\int d^4x \sqrt{-g}\left(-\frac{1}{4}g^{\mu\rho}g^{\nu\sigma}F_{\mu\nu}F_{\rho\sigma}\right), \quad F_{\mu\nu}=\partial_{\mu}A_{\nu}-\partial_{\nu}A_{\mu}.
\end{equation}
Note that for the purposes of the present work, in which we are interested only in couplings to gravity, we are neglecting the electromagnetic coupling between $\psi$ and $A_\mu$.

The next step is to linearize the metric around the Minkowski background. This leads to the standard Feynman rules describing the coupling of the graviton to the matter sector described above.

Tree-level exchanges of the graviton lead to the following effective  Lagrangian:
\begin{equation}
  \mathcal{L}_{\rm eff}=\frac{1}{M_P^2}\sum_i c_i \,O_i
\end{equation}
where the operators and their Wilson coefficients are given by \\ \\
Scalar operators:
\begin{flalign}
  \label{eq:scalar_vertex_first}
  &O_{\phi,1}=F(\phi)\Box F(\phi),\quad  c_{\phi,1}=-\frac{3}{4}\\
  &O_{\phi,2}= F(\phi)\partial_\mu\phi\,\partial^\mu\phi,\quad c_{\phi,2}= \frac{1}{2}\\
  &O_{\phi,3}= F(\phi)V(\phi),\quad c_{\phi,3}= -2\\
  &O_{\phi,4}= \partial^\mu\phi\,\partial^\nu\phi\frac{1}{\Box}\partial_\mu\phi\,\partial_\nu\phi,\quad c_{\phi,4}= \frac{1}{2}\\
  &O_{\phi,5}= \partial^\mu\phi\,\partial_\mu\phi\frac{1}{\Box}\partial^\nu\phi\,\partial_\nu\phi,\quad c_{\phi,5}=-\frac{1}{4}\\
  &O_{\phi,6}= \partial^\mu\phi\,\partial_\mu\phi\frac{1}{\Box}V(\phi),\quad c_{\phi,6}= 1\\
  \label{eq:scalar_vertex_last}
  &O_{\phi,7}= V(\phi)\frac{1}{\Box}V(\phi),\quad c_{\phi,7}=-1&&
\end{flalign}\\
Fermion operators:
\begin{flalign}
  &O_{\psi,1}=\left(i\bar{\psi}\,\gamma^{\mu}\overset{\leftrightarrow\hspace{2mm}}{\partial^{\nu}}\psi\right)\frac{1}{\Box}\left(i\bar{\psi}\,\gamma_{(\mu}\overset{\leftrightarrow}{\partial}_{\nu)}\psi\right), \quad c_{\psi,1}=\frac{1}{8}\\
  &O_{\psi,2}=m_\psi^2\, \bar\psi \psi\,\frac{1}{\Box}\,\bar\psi \psi, \quad c_{\psi,2}=-\frac{1}{4}&&
\end{flalign}\\
Vector operators:
\begin{flalign}
  &O_{A,1}=F_{\mu\rho}F^{\nu\rho}\frac{1}{\Box}F^{\mu\sigma}F_{\nu\sigma}, \quad c_{A,1}=\frac{1}{2}\\
  &O_{A,2}=F_{\mu\nu}F^{\mu\nu}\frac{1}{\Box}F_{\rho\sigma}F^{\rho\sigma}, \quad c_{A,2}=-\frac{1}{8}&&
\end{flalign}\\
Scalar-fermion operators:
\begin{flalign}
  &O_{\phi\psi,1}=m_\psi\, F(\phi)\bar\psi\psi, \quad c_{\phi\psi,1}=-\frac{1}{2}\\
  &O_{\phi\psi,2}=\partial^\mu\phi\,\partial^\nu\phi\,\frac{1}{\Box}\left(i\bar{\psi}\,\gamma_{\mu}\overset{\leftrightarrow}{\partial}_{\nu}\psi\right), \quad c_{\phi\psi,2}=\frac{1}{2}\\
  &O_{\phi\psi,3}=m_\psi\,V(\phi)\frac{1}{\Box}\, \bar\psi\psi, \quad c_{\phi\psi,3}=-1&&
\end{flalign}\\
Scalar-vector operators:
\begin{flalign}
  \label{eq:scal_vec_1}
  &O_{\phi A,1}=\partial^\mu\phi\,\partial_\nu\phi\,\frac{1}{\Box}F_{\mu\rho}F^{\nu\rho}, \quad c_{\phi A,1}=-1\\
  \label{eq:scal_vec_2}
  &O_{\phi A,2}=\partial^\mu\phi\,\partial_\mu\phi\,\frac{1}{\Box}F_{\nu\rho}F^{\nu\rho}, \quad c_{\phi A,2}=\frac{1}{4}&&
\end{flalign}\\
Fermion-vector operators:
\begin{flalign}
  &O_{\psi A,1}=\left(i\bar{\psi}\,\gamma^{\mu}\overset{\leftrightarrow}{\partial}_{\nu}\psi\right)\frac{1}{\Box}F_{\mu\rho}F^{\nu\rho}, \quad c_{\psi A,1}=-\frac{1}{2}\\
  \label{eq:fermion_vector_last}
  &O_{\psi A,2}=m_\psi\, \bar\psi\psi\,\frac{1}{\Box}F_{\nu\rho}F^{\nu\rho}, \quad c_{\psi A,2}=\frac{1}{4}&&
\end{flalign}\\
where the action of $\Box^{-1}$ on a function is given by the following convolution
\begin{equation}
  \label{eq:inverse_box}
  \frac{1}{\Box}f(x)=\int d^4y \,D(x-y) f(y),
\end{equation}
where $D(x-y)$ is the kernel
\begin{equation}
  \label{eq:integral_kernel}
  D(x-y)=\int\frac{d^4q}{(2\pi)^4}e^{iq\cdot(x-y)}\frac{-1}{q^2+i\varepsilon}.
\end{equation}
Note that as pointed out in \cite{Hill:2020oaj}, the effective operators arising from the non-minimal coupling to the Ricci scalar are completely local. We note that dimension 5 operators are not generated via tree-level exchange. All other operators are non-local and of dimension 6.  For further details on the derivation of these operators, we refer the reader to Appendix \ref{sec:appendixA}.

\section{Higher dimensional operators generated by graviton loops}
In this section, we study operators generated by graviton loops. Integrating out the quantum fluctuations of the metric leads to quantum gravitational corrections to the Hilbert-Einstein action which at second order in curvature read:
\begin{equation}\label{action1}
\begin{split}
  S= \int d^4x \, \sqrt{-g} &\left[  -\frac{1}{2}  M_P^2 \,\mathcal{R}+ c_1 \mathcal{R}^2 + c_2 \mathcal{R}_{\mu\nu}\mathcal{R}^{\mu\nu}   + b_1 \mathcal{R} \log \frac{\Box}{\mu^2}\mathcal{R} \right . \\
& \ \ \left . + b_2 \mathcal{R}_{\mu\nu}  \log \frac{\Box}{\mu^2}\mathcal{R}^{\mu\nu}  
+ b_3 \mathcal{R}_{\mu\nu\rho\sigma}  \log \frac{\Box}{\mu^2}\mathcal{R}^{\mu\nu\rho\sigma} 
   +\mathcal{O}(M_P^{-2})\right]+ S_{\text{matter}},
\end{split}
\end{equation}
where $\mathcal{R}$, $\mathcal{R}_{\mu\nu}$ and $\mathcal{R}_{\mu\nu\rho\sigma}$ are respectively the Ricci scalar, the Ricci tensor and the Riemann tensor.  The scale $\mu$ is the renormalization scale. See, e.g., \cite{Calmet:2019eof} for more details on the unique effective approach and further references.

It is easy to linearize this effective action to form the coupling between the new degrees of freedom $k^{\mu\nu}$ and $\sigma$ contained in the higher curvature terms and the energy momentum tensor. We find \cite{Calmet:2018hfb}
\begin{equation}
\begin{split}
  S=\int d^4 x \left[\left (- \frac{1}{2} h_{\mu\nu} \Box h^{\mu\nu}
 +\frac{1}{2} h_{\mu}^{\ \mu} \Box h_{\nu}^{\ \nu}  -h^{\mu\nu} \partial_\mu \partial_\nu h_{\alpha}^{\ \alpha}+ h^{\mu\nu} \partial_\rho \partial_\nu h^{\rho}_{\ \mu}\right) \right. \\ 
\left. + \left ( -\frac{1}{2} k_{\mu\nu} \Box k^{\mu\nu}
 +\frac{1}{2} k_{\mu}^{\ \mu} \Box k_{\nu}^{\ \nu}  -k^{\mu\nu} \partial_\mu \partial_\nu k_{\alpha}^{\ \alpha}+ k^{\mu\nu} \partial_\rho \partial_\nu k^{\rho}_{\ \mu}
 \right.  \right.  \\ \left. \left. 
 -\frac{m_2^2}{2} \left (k_{\mu\nu}k^{\mu\nu} - k_{\alpha}^{\ \alpha} k_{\beta}^{\ \beta} \right )
 \right)  \right.
  \\
 \left. + \frac{1}{2} \partial_\mu \sigma  \partial^\mu \sigma
  - \frac{m_0^2}{2} \sigma^2 - M_P \left(h_{\mu\nu}-k_{\mu\nu}+\frac{1}{\sqrt{3}} \sigma \eta_{\mu\nu}\right)T^{\mu\nu} 
  \right ].
\end{split}
\end{equation}
The masses of the spin-2 and spin-0 classical fields are given by \cite{Calmet:2018hfb}
\begin{align}
m_2^2=\frac{2}{ (b_2+ 4 b_3) \kappa^2 \,W\hspace{-0.8mm}\left(-\frac{2 \exp\frac{c_2}{(b_2+ 4 b_3)}}{ (b_2+ 4 b_3) \kappa^2 \mu^2}\right)},
\end{align}
\begin{align}
m_0^2=\frac{1}{ (3 b_1+b_2+ b_3) \kappa^2 \,W\hspace{-0.8mm}\left(-\frac{ \exp\frac{3 c_1+c_2}{(3 b_1+b_2+ b_3)}}{ (3 b_1+b_2+ b_3) \kappa^2 \mu^2}\right)},
\end{align}
and where $W(x)$ is the Lambert function.

We thus find that the only dimension five operators that can be generated by quantum gravity are those connecting the massive spin-0 and spin-2 to the energy-momentum tensor of the matter sector. In the limit of large masses $m_0$ and $m_2$, these fields generate local contact interactions, whereas the field $h_{\mu\nu}$ always generates non-local long-range interactions. By integrating out the extra gravitational degrees of freedom, one generates higher-dimensional operators suppressed by at least four powers of the Planck mass. Note that a dilaton-like field couples to the trace of the stress-energy tensor. This implies that the coupling of dilaton fields to two photons vanishes, however there are dimension five operators coupling Yang-Mills fields and the dilaton as the trace of the Yang-Mills stress-energy tensor does not vanish.

\section{Implications for ultralight dark matter models}

There has been much interest recently in hidden sector models for ultralight scalar dark matter models that could be probed with clocks or interferometers see e.g. \cite{Barontini:2021mvu} for a recent review. The coupling of these ultralight scalar fields is usually parametrized by dimension five or six operators normalized to the reduced Planck mass to the appropriate power. For example, their coupling to the photon is assumed to be of the form
\begin{eqnarray} \label{naiveops}
O_{dm,5}&=&\frac{d}{2 \sqrt{2} M_P}  \phi F_{\mu\nu}F^{\mu\nu}\\
O_{dm,6}&=&\frac{d}{8 M_P^2}  \phi^2 F_{\mu\nu}F^{\mu\nu}
\end{eqnarray}
where $d$ is some parameter of order unity.\footnote{ Atomic clocks can also constrain similar operators involving axion-like particles coupled to gluons, such as $a G_{\mu\nu}^b\tilde{G}^{b\mu\nu}$ (see \cite{Kim:2022ype,Sherrill:2023zah}).  However, the coupling of axion-like particles to photons, $a F_{\mu\nu}\tilde{F}^{\mu\nu}$, is predominantly probed using magnetometers. The same reasoning presented here for scalar fields also applies to pseudo-scalar ones.} One obvious question is how these operators can be generated in some fundamental theory incorporating both the standard model and dark matter. Our analysis demonstrates that these operators cannot be generated by quantum gravity, at least not at the perturbative level. 

Let us, however, now evaluate the operators $O_{\phi A,1}$ and $O_{\phi A,2}$ (eqs.~\eqref{eq:scal_vec_1} and \eqref{eq:scal_vec_2}) for when $\phi$ is given by a plane-wave solution as it is the case in these ultralight dark matter models
\begin{equation}
  \phi=\phi_0\cos(k\cdot x)
\end{equation}
where $k^\mu=(E,\vec{p}\,)$, we can calculate the quantity $\Box^{-1}\partial_\mu\phi\partial^\mu\phi$ in the following way:
\begin{equation}
  \partial_\mu\phi=k_\mu\,\phi_0\sin(k\cdot x) \ \  \implies \ \  \partial^\mu\phi\partial_\mu\phi=k^2\phi_0^2\sin^2(k\cdot x)=-\frac{1}{4}k^2\phi_0^2\left(e^{2ik\cdot x}+e^{-2ik\cdot x}-2\right)
\end{equation}
We can then perform a Fourier transformation and obtain
\begin{equation}
\begin{split}
  \mathcal{F}(\partial^\mu\phi\partial_\mu\phi)&=\int \frac{d^4y}{(2\pi)^4}e^{-iq\cdot y}\ \left(-\frac{1}{4}k^2\phi_0^2\left(e^{2ik\cdot y}+e^{-2ik\cdot y}-2\right)\right)=\\
  &=-\frac{k^2\phi_0^2}{4}\Bigl(\delta(q-2k)+\delta(q+2k)-2\delta(q)\Bigr).
\end{split}
\end{equation}
Now we can calculate 
\begin{equation}
\begin{split}
  \frac{1}{\Box}\partial_\mu\phi\partial^\mu\phi&=\int d^4q e^{iq\cdot x}\frac{-1}{q^2+i\varepsilon}\left(-\frac{k^2\phi_0^2}{4}\Bigl(\delta(q-2k)+\delta(q+2k)-2\delta(q)\Bigr)\right)=\\
  &=\frac{k^2\phi_0^2}{4}\left(\frac{e^{2ik\cdot x}}{4k^2}+\frac{e^{-2ik\cdot x}}{4k^2}\right)=\\
  &=\frac{\phi_0^2}{8}\cos(2k\cdot x)
\end{split}
\end{equation}
where we shifted the operator by a constant (infinite) quantity.

With an analogous calculation, we obtain 
\begin{equation}
  \frac{1}{\Box}\partial_\mu\phi\partial_\nu\phi=\frac{k_\mu k_\nu}{k^2}\frac{\phi_0^2}{8}\cos(2k\cdot x).
\end{equation}
If we consider now a stationary field, we have
\begin{equation}
  k^\mu=(E,\vec{0}\,)=(m_\phi,\vec{0}\,)
\end{equation}
Plugging this value of $k^\mu$ in the operators we get the results in \eqref{eq:scal_vec_1_new} and \eqref{eq:scal_vec_2_new}.

\begin{equation}
  \phi=\phi_0\cos(m_\phi t)
\end{equation}
Inserting back the factor $M_P^{-2}$ and the relevant Wilson coefficients,  we get
\begin{equation}
  \label{eq:scal_vec_1_new}
  O_{dm,1}=-\frac{\phi_0^2}{8 M_P^2}\cos(2m_\phi t) F_{0\mu}F^{0\mu}
\end{equation}
\begin{equation}
  \label{eq:scal_vec_2_new}
  O_{dm,2}=\frac{\phi_0^2}{32 M_P^2}\cos(2m_\phi t) F_{\mu\nu}F^{\mu\nu}.
\end{equation}
Note that this is precisely the form expected naively from the dimension 6 operator (\ref{naiveops}) if we insert a plane wave for the scalar field $\phi$. The dimension six operators naively considered in analyses of ultralight dark matter correspond to nonlocal operators generated by a graviton exchange.

Thus, the gravitationally generated interaction is always quadratic in the dark field, whereas many phenomenological analyses assume a linear coupling. Such linear couplings require either explicit UV interactions or non-perturbative gravitational physics.

\section{Global Symmetries at the Perturbative Level}

Consider an action that depends on matter fields $\chi$ and the metric $g_{\mu\nu}$, and is invariant under a generic transformation of the matter fields:
\begin{equation}
  \chi \mapsto \tilde{\chi}, \quad S[g_{\mu\nu},\chi] \mapsto S[g_{\mu\nu},\tilde{\chi}] = S[g_{\mu\nu},\chi].
\end{equation}
The stress-energy tensor,
\begin{equation}
  \label{eq:stress_energy_general}
  T_{\mu\nu} = \frac{2}{\sqrt{-g}} \frac{\delta S_{\rm matter}}{\delta g^{\mu\nu}},
\end{equation}
is then also invariant under this transformation. This follows from the decomposition in eq.~\eqref{eq:action_decomposition}, which implies that $S_{\rm matter}$ is invariant (since $S_{\rm EH}$ depends only on $g_{\mu\nu}$ and is itself invariant). Consequently, the right-hand side of eq.~\eqref{eq:stress_energy_general} is invariant, making $T_{\mu\nu}$ an invariant quantity.

The tree-level effective action therefore inherits this invariance under transformations of $\chi$, as it depends solely on $T_{\mu\nu}$ (see eq.~\eqref{eq:effective_action_general}). Thus, any global symmetry of the matter fields in the original action $S$ is preserved in the tree-level effective action $S_{\rm eff}$.

It is often argued that a complete theory of quantum gravity must break all global symmetries (see \cite{Banks:1988yz}; for a modern review of the swampland conjectures, see e.g.~\cite{Grana:2021zvf}). However, the analysis above demonstrates that this does not occur at the perturbative tree level. For instance, in the Standard Model, baryon and lepton number remain conserved under tree-level gravitational effects. Processes such as proton decay mediated by gravity would therefore require highly non-perturbative contributions or the explicit inclusion of proton-decay operators in the original action $S$, as in grand unified theories (see e.g.~\cite{Barr:2012xb}).

\section{Conclusions}
We have considered the standard model coupled gravitationally to some hidden sector containing fields of different spins. We have calculated the leading-order effective operators generated in perturbative quantum gravity at tree-level and at the one-loop level. One main result is that dimension five operators coupling standard model fields and those of the hidden sector are not generated. The leading order operators are non-local and of dimension 6. As expected, loops generate operators of dimension 8. When applied to plane wave configurations of ultralight dark matter, we recover interactions of the form arising from local dimension 6 operators previously discussed in the literature.  Furthermore, we argued that the tree-level symmetries of the original action are preserved in the quantum gravitational effective action.  Consequently, perturbative quantum gravity does not induce symmetry-violating operators and cannot mediate processes such as proton decay.

Finally, we have only considered effective operators arising from small graviton fluctuations (i.e., perturbation theory). Strong fluctuations in spacetime at the Planck scale could manifest as operators whose action resembles that of black hole absorption and emission of particles. These non-perturbative effects may lead to a dimension five operator in which, e.g., a scalar is absorbed by a small quantum black hole and two photons are emitted in its evaporation. (Or, alternatively, the time-reversed process.) The quantum black hole has a minimum mass of order the Planck scale, so it does not appear explicitly in the low-energy effective theory, although it can induce operators involving the light fields. Detection of this dimension five interaction might be direct evidence of strong field fluctuations in spacetime due to quantum gravity.

{\it Acknowledgments:}
	 The work of T.A. is supported by a doctoral studentship of the Science and Technology Facilities Council (training grant No. ST/Y509620/1, project ref. 2917813).
\\
\\
{\it Data Availability Statement:}
	This manuscript has no associated data. Data sharing not applicable to this article as no datasets were generated or analysed during the current study.
	
\newpage

\appendix
\section{Calculation of the operators}
\label{sec:appendixA}

To calculate the effective action arising from the contribution of the tree-level diagrams with a graviton mediator we will follow the approach in \cite{Hill:2020oaj}. We can expand the metric around a Minkowski background:
\begin{equation}
  g_{\mu\nu}=\eta_{\mu\nu}+\frac{1}{M_P}h_{\mu\nu},
\end{equation}
and all the other quantities depending on the metric are expanded accordingly. The Einstein-Hilbert action upon this expansion becomes the Fierz-Pauli action:
\begin{equation}
  S_{EH}=\frac{1}{8}\int d^4x\left(h\Box h-h^{\mu\nu}\Box h_{\mu\nu}-2h\partial_\mu\partial_\nu h^{\mu\nu}+2h^{\mu\nu}\partial_\mu\partial^\rho h_{\nu\rho}\right).
\end{equation}
To define the graviton propagator we impose the de Donder gauge condition:
\begin{equation}
  \partial_\mu h^{\mu\nu}=\frac{1}{2}\partial^\nu h.
\end{equation}
Now the Fierz-Pauli action reduces to:
\begin{equation}
  S_{\rm EH}=\frac{1}{2}\int d^4x\ h^{\mu\nu}\left(\frac{1}{8}P_{\mu\nu\rho\sigma}\right)\Box h^{\rho\sigma}, \quad P_{\mu\nu\rho\sigma}=\eta_{\mu\rho}\eta_{\nu\sigma}+\eta_{\mu\sigma}\eta_{\nu\rho}-\eta_{\mu\nu}\eta_{\rho\sigma}.
\end{equation}
Inverting the kinetic operator we can find the form of the free graviton propagator as
\begin{equation}
  \bra{0}\hat{T}\,h_{\mu\nu}(x)\,h_{\rho\sigma}(y)\ket{0}=2i\,P_{\mu\nu\rho\sigma} \,D(x-y),
\end{equation}
where we introduced the kernel $D(x-y)$ as in eq.~\eqref{eq:integral_kernel}, which satisfies
\begin{equation}
  \Box_x\, D(x-y)=\delta^4(x-y).
\end{equation}
Now the matter action can be decomposed as a sum of the action of Minkowski background and a coupling to gravity:
\begin{equation}
  S_{\rm matter}[g_{\mu\nu}]=S_{\rm matter}[\eta_{\mu\nu}]+S_{\rm 1},
\end{equation}
where the coupling is proportional to the stress-energy tensor:
\begin{equation}
  S_1=-\frac{1}{2M_P}\int d^4x\ h^{\mu\nu}\,T_{\mu\nu},
\end{equation}
where the stress-energy tensor is evaluated at the Minkowski background. We can now easily calculate the tree-level contributions as present them in the form of a (non-local) effective action, and the result is
\begin{equation}
\begin{split}
  \label{eq:effective_action_general}
  S_{\rm eff}&=\frac{1}{4M_P^2}\int d^4x\int d^4y \,T^{\mu\nu}(x) \ P_{\mu\nu\rho\sigma}\, D(x-y)\ T^{\rho\sigma}(y)=\\
  &=\frac{1}{4M_P^2}\int d^4x\left(2\,T^{\mu\nu}\frac{1}{\Box}T_{\mu\nu}-T\frac{1}{\Box}T\right),
\end{split}
\end{equation}
where we defined $\Box^{-1}$ acting on a function to be the convolution given in eq.~\eqref{eq:inverse_box}.

From eq.~\eqref{eq:matter_action}, the stress-energy tensor of the theory can be decomposed as such
\begin{equation}
  T_{\mu\nu}=\frac{2}{\sqrt{-g}}\frac{\delta S_{\rm matter}}{\delta g^{\mu\nu}}=T^\phi_{\mu\nu}+T^{\psi}_{\mu\nu}+T^{A}_{\mu\nu}.
\end{equation}
The scalar part is further decomposed as
\begin{equation}
  T^\phi_{\mu\nu}=T^{\phi,1}_{\mu\nu}+T^{\phi,2}_{\mu\nu},
\end{equation}
where $T^{\phi,1}_{\mu\nu}$ represents the stress-energy tensor coming from the minimal-coupling sector
\begin{equation}
  T^{\phi,1}_{\mu\nu}= \partial_\mu\phi\,\partial_\nu\phi-g_{\mu\nu}\left(\frac{1}{2}g^{\rho\sigma}\partial_\rho\phi\,\partial_\sigma\phi-V(\phi)\right),
\end{equation}
while $T^{\phi,2}_{\mu\nu}$ represents the stress-energy tensor of the non-minimal coupling:
\begin{equation}
T^{\phi,2}_{\mu\nu}=\left(G_{\mu\nu}+g_{\mu\nu}\,g^{\rho\sigma}\nabla_\rho\nabla_\sigma-\nabla_\mu\nabla_\nu\right)F(\phi).
\end{equation}
The stress-energy tensor of the fermion field is
\begin{equation}
  T^\psi_{\mu\nu}= \frac{i}{2}\bar{\psi}\left(\gamma_{(\mu}\overset{\rightarrow}{D}_{\nu)}-\overset{\leftarrow}{D}_{(\mu}\gamma_{\nu)}\right)\psi
\end{equation}
The stress-energy tensor of the vector field is
\begin{equation}
  T^A_{\mu\nu}=-g^{\rho\sigma}F_{\mu\rho}F_{\nu\sigma}+\frac{1}{4}g_{\mu\nu}g^{\rho\sigma}g^{\lambda\tau}F_{\rho\lambda}F_{\sigma\tau}.
\end{equation}

Since $S_{\rm eff}$ is quadratic in $T_{\mu\nu}$, it can be expanded as the following sum
\begin{equation}
  S_{\rm eff}=S_{\phi\phi}+S_{\phi'\phi'}+S_{\psi\psi}+S_{AA}+S_{\phi\phi'}+S_{\phi\psi}+S_{\phi A}+S_{\phi' \psi}+S_{\phi' A}+S_{\psi A},
\end{equation}
where the subscripts $\phi$, $\phi'$, $\psi$ and $A$ denote contributions from the tensors $T^{\phi,1}_{\mu\nu}$, $T^{\phi,2}_{\mu\nu}$, $T^{\psi}_{\mu\nu}$ and $T^{A}_{\mu\nu}$ respectively, which are now evaluated on the Minkowski background.

Now, as pointed out in \cite{Hill:2020oaj}, the sector of $S_{\rm eff}$ arising from the non-minimal coupling in $T^{\phi,2}_{\mu\nu}$ is completely local. \\
$\phi'\phi'$ sector:
\begin{equation}
  S_{\phi'\phi'}=-\frac{3}{4M_P^2}\int d^4x \ F(\phi)\,\Box F(\phi)
\end{equation}
$\phi\phi'$ sector:
\begin{equation}
  S_{\phi\phi'}=-\frac{1}{2M_P^2}\int d^4x\ F(\phi)\,T^{\phi,1}=\frac{1}{2M_P^2}\int d^4x\Bigl(F(\phi)\partial_\mu\phi\,\partial^\mu\phi-4F(\phi)V(\phi)\Bigr)
\end{equation}
$\phi'\psi$ sector:
\begin{equation}
  S_{\phi'\psi}=-\frac{1}{2M_P^2}\int d^4x\ F(\phi)\,T^{\psi}=-\frac{1}{4M_P^2}\int d^4x\ F(\phi)\left(i\bar\psi\,\overset{\leftrightarrow}{\slashed{\partial}}\,\psi\right)
\end{equation}
$\phi'A$ sector is not present, since $T_{\mu\nu}^A$ is traceless:
\begin{equation}
  S_{\phi'A}=-\frac{1}{2M_P^2}\int d^4x\ F(\phi)\,T^{A}=0
\end{equation}
$\phi\phi$ sector:
\begin{equation}
\begin{split}
  S_{\phi\phi}&=\frac{1}{4M_P^2}\int d^4x\left(2\,T^{\phi,1\, \mu\nu}\frac{1}{\Box}T^{\phi,1}_{\mu\nu}-T^{\phi,1}\frac{1}{\Box}T^{\phi,1}\right)=\\
  &=\frac{1}{4M_P^2}\int d^4x\biggl(2\partial^\mu\phi\,\partial^\nu\phi\frac{1}{\Box}\partial_\mu\phi\,\partial_\nu\phi-\partial^\mu\phi\,\partial_\mu\phi\frac{1}{\Box}\partial^\nu\phi\,\partial_\nu\phi\,+\\
  & \hspace{3.5cm}+4\partial^\mu\phi\,\partial_\mu\phi\frac{1}{\Box}V(\phi)-4V(\phi)\frac{1}{\Box}V(\phi)\biggr)
\end{split}
\end{equation}
$\psi\psi$ sector:
\begin{equation}
\begin{split}
  S_{\psi\psi}&=\frac{1}{4M_P^2}\int d^4x\left(2\,T^{\psi\, \mu\nu}\frac{1}{\Box}T^\psi_{\mu\nu}-T^{\psi}\frac{1}{\Box}T^\psi\right)=\\
  &=\frac{1}{16M_P^2}\int d^4x\biggl(2\left(i\bar{\psi}\,\gamma^{\mu}\overset{\leftrightarrow\hspace{2mm}}{\partial^{\nu}}\psi\right)\frac{1}{\Box}\left(i\bar{\psi}\,\gamma_{(\mu}\overset{\leftrightarrow}{\partial}_{\nu)}\psi\right)-\left(i\bar\psi \overset{\leftrightarrow}{\slashed{\partial}}\psi\right)\frac{1}{\Box}\left(i\bar\psi \overset{\leftrightarrow}{\slashed{\partial}}\psi\right)\biggr)
\end{split}
\end{equation}
$AA$ sector:
\begin{equation}
\begin{split}
  S_{AA}&=\frac{1}{2M_P^2}\int d^4x\left(T^{A\, \mu\nu}\frac{1}{\Box}T^A_{\mu\nu}\right)=\\
  &=\frac{1}{8M_P^2}\int d^4x\left(4F_{\mu\rho}F^{\nu\rho}\frac{1}{\Box}F^{\mu\sigma}F_{\nu\sigma}-F_{\mu\nu}F^{\mu\nu}\frac{1}{\Box}F_{\rho\sigma}F^{\rho\sigma}\right)
\end{split}
\end{equation}
$\phi\psi$ sector:
\begin{equation}
\begin{split}
  S_{\phi \psi}&=\frac{1}{2M_P^2}\int d^4x\left(2\,T^{\phi,1\, \mu\nu}\frac{1}{\Box}T^\psi_{\mu\nu}-T^{\phi,1}\frac{1}{\Box}T^\psi\right)=\\
  &=\frac{1}{2M_P^2}\int d^4x\biggl(\partial^\mu\phi\,\partial^\nu\phi\,\frac{1}{\Box}\left(i\bar{\psi}\,\gamma_{\mu}\overset{\leftrightarrow}{\partial}_{\nu}\psi\right)-V(\phi)\frac{1}{\Box} \left(i\bar\psi \overset{\leftrightarrow}{\slashed{\partial}}\psi\right)\biggr)
\end{split}
\end{equation}
$\phi A$ sector:
\begin{equation}
\begin{split}
  S_{\phi A}&=\frac{1}{M_P^2}\int d^4x\left(T^{\phi,1\, \mu\nu}\frac{1}{\Box}T^A_{\mu\nu}\right)=\\
  &=\frac{1}{4M_P^2}\int d^4x\left(-4\partial^\mu\phi\,\partial_\nu\phi\,\frac{1}{\Box}F_{\mu\rho}F^{\nu\rho}+\partial^\mu\phi\,\partial_\mu\phi\,\frac{1}{\Box}F_{\nu\rho}F^{\nu\rho}\right)
\end{split}
\end{equation}
$\psi A$ sector:
\begin{equation}
\begin{split}
  S_{\psi A}&=\frac{1}{M_P^2}\int d^4x\left(T^{\psi\, \mu\nu}\frac{1}{\Box}T^A_{\mu\nu}\right)=\\
  &=\frac{1}{8M_P^2}\int d^4x\left(-4\left(i\bar{\psi}\,\gamma^{\mu}\overset{\leftrightarrow}{\partial}_{\nu}\psi\right)\frac{1}{\Box}F_{\mu\rho}F^{\nu\rho}+\left(i\bar\psi \overset{\leftrightarrow}{\slashed{\partial}}\psi\right)\frac{1}{\Box}F_{\nu\rho}F^{\nu\rho}\right)
\end{split}
\end{equation}
By collecting the operators in this action, we arrive at the form presented in eqs.~\eqref{eq:scalar_vertex_first} to \eqref{eq:fermion_vector_last}.


\end{document}